\documentclass[aps,prd,preprint,tightenlines,groupedaddress,showpacs]{revtex4}
\usepackage{epsf,epsfig,graphics,graphicx,amssymb,amsthm}
\begin{document}

\title{Inflaton fluctuations in the presence of cosmological defects}

\author{Hing-Tong Cho$^1$}
\author{Kin-Wang Ng$^2$}
\author{I-Chin Wang$^1$}
\affiliation{
$^1$Department of Physics, Tamkang University, Tamsui, New Taipei City 25137, Taiwan\\
$^2$Institute of Physics, Academia Sinica, Taipei 11529, Taiwan}

\begin{abstract}
We study quantum fluctuations of a free massless scalar field during inflation
in the presence of a point, line, or plane defect such as a black hole, cosmic string, or domain wall,
using a perturbative expansion in powers of small defect parameters.
We provide results for the scalar two-point correlation functions that show explicitly a
small violation of translational invariance during inflation.
\end{abstract}

\maketitle

\date{\today}

\section{Introduction}

The cosmic microwave background (CMB) that we observe today is almost
homogeneous and isotropic. The background temperature in our sky is
about $2.7$K with tiny fluctuations at a level of about
$10^{-5}$K. Measurements of the power spectrum of these fluctuations,
combined with other observations of the large-scale sturctures of the Universe,
concordantly prevail a spatially flat universe~\cite{cosmology}.
Inflation scenario offers compelling explanations for the homogeneity, the
isotropy, and the flatness of the present Universe~\cite{olive}.
Moreover, quantum fluctuations of the inflaton field during a slow-roll
inflation give rise to primordial density fluctuations with a nearly
scale-invariant power spectrum, in good agreement with the recent
Planck data on CMB anisotropies~\cite{planckXVI}.

However, there have been some hints of new cosmological physics beyond the
standard slow-roll inflation model. For examples, the WMAP and Planck CMB data show the so-called
large-scale anomalies, such as the low quadrupole, a hemispherical power asymmetry,
and a remarkable alignment of the quadrupole and octupole, although
the evidences for these anomalies are not statistically significant~\cite{wmap7,wmap9ng,planckXXIII}.
In addition, the combined Planck, BICEP2, and large-scale-structure data prefer, though not yet confirmed,
a slightly negative running in the spectral index~\cite{planckXVI,bicep2},
which is unexpected in the standard slow-roll inflation model.
Future CMB data will soon confirm or set tighter constraints on these possibilities.

Recently, the authors in Ref.~\cite{carroll} proposed potentially observable effects
of a small violation of translational invariance during inflation, as
characterized by the presence of a preferred point, line, or plane.
The violation may induce deviations from pure statistical isotropy of
cosmological perturbations, thus leaving anomalous imprints on the
CMB anisotropy~\cite{carroll}.
In this paper, we will provide a mechanism for generating
a violation of translational invariance during inflation in which
the inflaton fluctuates in the presence of cosmological defects such as monopoles or black holes,
cosmic strings, and domain walls. This is equivalent to considering quantum fluctuations of
a free massless scalar field in a background metric pertinent to the presence of
such cosmological defects in the de Sitter space.
Since the violation is expected to be small, instead of pursuing exact solutions of
the scalar field equation in such background metrics, we will consider small defect
parameters and develop a perturbation approach to tackle the problem.

Here we simply assume the presence of cosmological defects in inflation and consider their
effects on inflaton quantum fluctuations. The production mechanism for these defects
is not the main interest of the present paper and will be briefly addressed.
The paper is organized as follows. In Sec.~\ref{pert}, we will lay out the perturbation method.
Then, it will be applied in Sec.~\ref{defect} to calculate the scalar quantum fluctuations due to
the cosmological defects. Sec.~\ref{conclusion} contains our conclusions.
In Appendix A, we will use the in-in formalism to re-derive the corresponding quantum fluctuations
due to defects and to compare with the results obtained in previous sections.

\section{Perturbation method}
\label{pert}

With these considerations in mind, in this work we will
investigate the quantum fluctuations of a free massless scalar field in the
metric that takes the form,
\begin{equation}
ds^2=g_{\mu\nu} dx^\mu dx^\nu,\quad x^\mu=(\tau,\vec{x}),
\label{firstmetric}
\end{equation}
where $d\tau=a^{-1}(\tau)dt$ is the conformal time, the cosmic scale factor $a(t)=e^{Ht}$
with $H$ being the Hubble parameter in inflation, and
\begin{equation}
g_{\mu\nu}=g^{(0)}_{\mu\nu}+g^{(1)}_{\mu\nu},\quad g^{(0)}_{\mu\nu}={\rm diag}(-a^{2},a^{2},a^{2},a^{2}).
\label{pertmetric}
\end{equation}
Here $g^{(1)}_{\mu\nu}$ can be arbitrarily small. Then, the contravariance and trace of the metric can be expanded as
\begin{equation}
g^{\mu\nu}=g^{(0)\mu\nu}-g^{(0)\mu\alpha}g^{(1)}_{\alpha\beta}g^{(0)\beta\nu}+\cdots\,\,\,,
\label{contra}
\end{equation}
\begin{equation}
\sqrt{-g}=\sqrt{-g^{(0)}}+\frac{1}{2}\sqrt{-g^{(0)}}g^{(0)\mu\nu}g^{(1)}_{\mu\nu}+\cdots\,\,\,.
\label{trace}
\end{equation}
Here we consider the quantized scalar field,
\begin{equation}
\hat{\phi}(x)=\int d^{3}k
\left[\hat{a}_{\vec{k}}\varphi_{\vec{k}}(x)
+\hat{a}^{\dagger}_{\vec{k}}\varphi_{\vec{k}}^{*}(x)\right],
\end{equation}
with the commutation relations given by
\begin{eqnarray}
&&[\hat{a}_{\vec{k}},\hat{a}_{\vec{k}'}]=[\hat{a}^{\dagger}_{\vec{k}},\hat{a}^{\dagger}_{\vec{k}'}]=0,\nonumber\\
&&[\hat{a}_{\vec{k}},\hat{a}^{\dagger}_{\vec{k}'}]=\delta(\vec{k}-\vec{k}').
\end{eqnarray}
The vacuum state is defined as
\begin{equation}
\hat{a}_{\vec{k}}|0\rangle=0.
\end{equation}

The mode function $\varphi_{\vec{k}}$  satisfies the Klein-Gordon equation,
\begin{equation}
\partial_{\mu}\left(\sqrt{-g}g^{\mu\nu}\partial_{\nu}\varphi_{\vec{k}}\right)=0.
\end{equation}
Expanding $\varphi_{\vec{k}}=\varphi_{\vec{k}}^{(0)}+\varphi_{\vec{k}}^{(1)}+\cdots$
and using Eqs.~(\ref{contra}) and~(\ref{trace}), this equation becomes
\begin{equation}
\partial_{\mu}\left[\sqrt{-g^{(0)}}\left(1+\frac{1}{2}g^{(0)\rho\sigma}g^{(1)}_{\rho\sigma}\right)
\left(g^{(0)\mu\nu}-g^{(0)\mu\alpha}g^{(0)\nu\beta}g^{(1)}_{\alpha\beta}\right)\partial_{\nu}
\left(\varphi^{(0)}_{\vec{k}}+\varphi^{(1)}_{\vec{k}}+\cdots\right)\right]=0,
\end{equation}
To the zeroth order, we have the homogeneous equation,
\begin{equation}
\partial_{\mu}\left(\sqrt{-g^{(0)}}g^{(0)\mu\nu}\partial_{\nu}\varphi^{(0)}_{\vec{k}}\right)=0\,,
\end{equation}
with a homogeneous solution, that is, choosing the Bunch-Davies vacuum,
\begin{equation}
\varphi^{(0)}_{\vec{k}}(\tau,\vec{x})=-\frac{H}{2^{5/2}\pi
k^{3/2}}(-k\tau)^{3/2}H^{(1)}_{3/2}(-k\tau)e^{i\vec{k}\cdot\vec{x}}\,.
\label{homo}
\end{equation}
The first order equation is given by
\begin{equation}
\partial_{\mu}\left(\sqrt{-g^{(0)}}g^{(0)\mu\nu}\partial_{\nu}\varphi^{(1)}_{\vec{k}}\right)=J_{\vec{k}}
\end{equation}
where
\begin{equation}
J_{\vec{k}}=\partial_{\mu}\left[\sqrt{-g^{(0)}}
\left(g^{(0)\mu\alpha}g^{(0)\nu\beta}g^{(1)}_{\alpha\beta}-\frac{1}{2}g^{(0)\rho\sigma}g^{(1)}_{\rho\sigma}g^{(0)\mu\nu}\right)
\partial_{\nu}\varphi^{(0)}_{\vec{k}}\right]\,.\label{source}
\end{equation}
In Cartesian coordinates, the first order equation becomes
\begin{equation}
\left(\partial^{2}_{\tau}-\frac{2}{\tau}\partial_{\tau}-\vec{\nabla}^{2}\right)\varphi^{(1)}_{\vec{k}}=-\frac{1}{a^{2}}J_{\vec{k}}\,.
\end{equation}
To solve this equation, we may consider the Green's function that satisfies
\begin{equation}
\left(\partial^{2}_{\tau}-\frac{2}{\tau}\partial_{\tau}-\vec{\nabla}^{2}\right)G(x,x')=\delta(\tau-\tau')
\delta(\vec{x}-\vec{x}')\,.
\end{equation}
Suppose that
\begin{equation}
G(x,x')=\int d^{3}k
g_{\vec{k}}(\tau,\tau')e^{i\vec{k}\cdot(\vec{x}-\vec{x}')},
\end{equation}
then we have
\begin{equation}
\left(\partial^{2}_{\tau}-\frac{2}{\tau}\partial_{\tau}+k^{2}\right)g_{\vec{k}}(\tau,\tau')=
\frac{1}{(2\pi)^{3}}\delta(\tau-\tau')\,.
\end{equation}
For the retarded Green's function, $g_{k}(\tau,\tau')=0$ for $\tau'>\tau>\tau_{i}$,
where $\tau_i$ denotes an initial time when the source begins to
operate. For $0>\tau>\tau'$,
\begin{eqnarray}
g_{\vec{k}}(\tau,\tau')&=&\frac{i}{32\pi^{2}\tau'^{2}k^{3}}\left[(-k\tau)^{3/2}H^{(1)}_{3/2}(-k\tau)
(-k\tau')^{3/2}H^{(2)}_{3/2}(-k\tau')\right.\nonumber\\
 & &\,\,\,\,\,\,\,\,\,\,\,\,\,\,\,\,\,\,\,\,\,\,\,\,\,\,\left.-(-k\tau')^{3/2}H^{(1)}_{3/2}(-k\tau')
(-k\tau)^{3/2}H^{(2)}_{3/2}(-k\tau)\right].
\label{retardg}
\end{eqnarray}
Hence, we obtain the first order correction for the mode function,
\begin{eqnarray}
\varphi^{(1)}_{\vec{k}}(\tau,\vec{x})& = & -\int^{0}_{\tau_{i}}d\tau'\int
d^{3}x'G(\tau,\vec{x};\tau',\vec{x}')
J_{\vec{k}}(\tau',\vec{x}')\frac{1}{a^{2}(\tau')}\nonumber\\
& = & -\int^{\tau}_{\tau_{i}}d\tau'\int d^{3}x'\int d^{3}k'
g_{\vec{k}'}(\tau,\tau')e^{i\vec{k}'\cdot(\vec{x}-\vec{x}')}
J_{\vec{k}}(\tau',\vec{x}')\frac{1}{a^{2}(\tau')}\,\,\,,
\label{phi1k}
\end{eqnarray}
which can be written in the form,
\begin{equation}
\varphi^{(1)}_{\vec{k}}(\tau,\vec{x})=\epsilon\int d^{3}k'
\left[\alpha_{\vec{k},\vec{k}'}(\tau)\varphi^{(0)}_{\vec{k}'}(\tau,\vec{x})
+\beta_{\vec{k},\vec{k}'}(\tau)\varphi^{(0)\ast}_{\vec{k}'}(\tau,\vec{x})\right]\,,
\label{phione}
\end{equation}
where $\epsilon$ is a real expansion parameter (see below for details). After inserting
Eq.~(\ref{retardg}) into Eq.~(\ref{phi1k}) and using the homogeneous solution~(\ref{homo}),
$\alpha_{\vec{k},\vec{k}'}$ and $\beta_{\vec{k},\vec{k}'}$ can be
written as
\begin{eqnarray}
\epsilon\alpha_{\vec{k},\vec{k}'}(\tau)&=&\frac{i H}{2^{5/2}\pi
k'^{3/2}}\int_{\tau_{i}}^{\tau}d\tau'\int d^{3}x'
J_{\vec{k}}(\tau',\vec{x}')(-k'\tau')^{3/2}H^{(2)}_{3/2}(-k'\tau')e^{-i\vec{k}'\cdot\vec{x}'}\,\,\,,\label{alpha}
\end{eqnarray}
\begin{eqnarray}
\epsilon\beta_{\vec{k},\vec{k}'}(\tau)&=&-\frac{i H}{2^{5/2}\pi
k'^{3/2}}\int_{\tau_{i}}^{\tau}d\tau'\int d^{3}x'
J_{\vec{k}}(\tau',\vec{x}')(-k'\tau')^{3/2}H^{(1)}_{3/2}(-k'\tau')e^{i\vec{k}'\cdot\vec{x}'}\,\,\,.\label{beta}
\end{eqnarray}

The two-point correlation function is then given by
\begin{eqnarray}
\Delta(\vec{x},\vec{x}')&\equiv&
\langle \hat{\phi}(\tau,\vec{x})\hat{\phi}(\tau,\vec{x}') \rangle \\&=& \int d^{3}k
\varphi_{\vec{k}}^{(0)}(\tau,\vec{x})\varphi_{\vec{k}}^{(0)\ast}(\tau,\vec{x}')\nonumber\\&&
+\int
d^{3}k\left[\varphi_{\vec{k}}^{(0)}(\tau,\vec{x})\varphi_{\vec{k}}^{(1)\ast}(\tau,\vec{x}')+
\varphi_{\vec{k}}^{(0)\ast}(\tau,\vec{x}')\varphi_{\vec{k}}^{(1)}(\tau,\vec{x})\right]+ \cdots\,\,.\label{twopointcorrelator}
\end{eqnarray}
The first term on the right hand side of the above equation is the zeroth order correlation,
\begin{equation}
\Delta^{(0)}(\vec{x},\vec{x}')=\int d^{3}k
\varphi_{\vec{k}}^{(0)}(\tau,\vec{x})\varphi_{\vec{k}}^{(0)\ast}(\tau,\vec{x}')
=\frac{H^2}{4\pi^2}\int d^{3}k \frac{1}{4\pi k^3} (1+k^2\tau^2)e^{i\vec{k}\cdot( \vec{x}-\vec{x}')} ,
\end{equation}
which reproduces the well-known scale-invariant power spectrum of de Sitter quantum fluctuations~\cite{desitter}.
The second term is the first order correction,
\begin{eqnarray}
\Delta^{(1)}(\vec{x},\vec{x}')&=&\int
d^{3}k\left[\varphi_{\vec{k}}^{(0)}(\tau,\vec{x})\varphi_{\vec{k}}^{(1)\ast}(\tau,\vec{x}')+
\varphi_{\vec{k}}^{(0)\ast}(\tau,\vec{x}')\varphi_{\vec{k}}^{(1)}(\tau,\vec{x})\right]\nonumber\\
&=&\epsilon\int d^{3}k\left\{ \varphi_{\vec{k}}^{(0)}(\tau,\vec{x})\int
d^{3}k'\left[
\alpha_{\vec{k},\vec{k}'}^{\ast}(\tau)\varphi^{(0)\ast}_{\vec{k}'}(\tau,\vec{x}')
+\beta_{\vec{k},\vec{k}'}^{\ast}(\tau)\varphi^{(0)}_{\vec{k}'}(\tau,\vec{x}')\right]\right.\nonumber\\
& &\,\,\,\,\,\,\,\,\,\,\,\,
+\left.\varphi_{\vec{k}}^{(0)\ast}(\tau,\vec{x}')\int d^{3}k'\left[
\alpha_{\vec{k},\vec{k}'}(\tau)\varphi^{(0)}_{\vec{k}'}(\tau,\vec{x})
+\beta_{\vec{k},\vec{k}'}(\tau)\varphi^{(0)\ast}_{\vec{k}'}(\tau,\vec{x})\right]\right\}\nonumber\\
&=&\epsilon\int d^{3}k d^{3}k'
e^{i\vec{k}\cdot\vec{x}+i\vec{k}'\cdot\vec{x}'}
\left[\alpha_{\vec{k},-\vec{k}'}^{\ast}(\tau)\varphi_{k}^{(0)}(\tau)\varphi_{k'}^{(0)\ast}(\tau)+
\beta_{\vec{k},\vec{k}'}^{\ast}(\tau)\varphi_{k}^{(0)}(\tau)\varphi_{k'}^{(0)}(\tau)\right.\nonumber\\
& & \,\,\,\,\,\,\,\,\,\,\,\,\,\,\,\,\,\,\,\,\,\,\,\,\,\,\,\,\,\,
+\left.\alpha_{-\vec{k}',\vec{k}}(\tau)\varphi_{k}^{(0)}(\tau)\varphi_{k'}^{(0)\ast}(\tau)+
\beta_{-\vec{k}',-\vec{k}}(\tau)\varphi_{k}^{(0)\ast}(\tau)\varphi_{k'}^{(0)\ast}(\tau)\right]\,\,,
\end{eqnarray}
where we have used Eq.~(\ref{phione}) and written $\varphi^{(0)}_{\vec{k}}(\tau,\vec{x})=\varphi_{k}(\tau)e^{i\vec{k}\cdot\vec{x}}$.
Thus, we define the power spectrum of the first order correction as
\begin{eqnarray}
\Delta^{(1)}(\vec{k},\vec{k'})&=&\epsilon\left[\alpha_{\vec{k},-\vec{k}'}^{\ast}(\tau)\varphi_{k}^{(0)}(\tau)\varphi_{k'}^{(0)\ast}(\tau)+
\beta_{\vec{k},\vec{k}'}^{\ast}(\tau)\varphi_{k}^{(0)}(\tau)\varphi_{k'}^{(0)}(\tau)\right.\nonumber\\
&  & +
\left.\alpha_{-\vec{k}',\vec{k}}(\tau)\varphi_{k}^{(0)}(\tau)\varphi_{k'}^{(0)\ast}(\tau)+
\beta_{-\vec{k}',-\vec{k}}(\tau)\varphi_{k}^{(0)\ast}(\tau)\varphi_{k'}^{(0)\ast}(\tau)\right]\label{kspectrum}\,\,\,.
\end{eqnarray}
In the next section, we will consider some explicit examples that the metrics are not
homogeneous and isotropic, and will derive respectively the first order
corrections of the inflaton two-point correlation function.

\section{Defects in inflation}
\label{defect}

The cosmic no hair conjecture infers that the inflationary
universe approaches asymptotically the de Sitter spacetime till
the end of inflation~\cite{hawking77}.  Nevertheless, the effects of
matter and spacetime inhomogeneities on inflation should be
considered as long as the duration of inflation is finite, our Universe
accidentally locates in the vicinity of these inhomogeneities,
or inhomogeneities are being produced during inflation.

If inflation is a phase transition such as a symmetry
breaking process, topological defects such as monopoles, strings, or
domain walls will be formed~\cite{symmetry} during the transition.
Several authors have studied the onset of inflation
under inhomogeneous initial conditions to determine whether large
inhomogeneity during the very early Universe can prevent the
Universe from entering an inflationary era~\cite{inhom}. It was
found that in some cases a large initial inhomogeneity may
suppress the onset of inflation~\cite{gold}. If the inflaton field
is sufficiently inhomogeneous, the wormhole can form from collapsing
vacuum energy density peaks before the inhomogeneity is damped by the exponential
expansion~\cite{holcomb}. In the case of inhomogeneities in a dust
era before inflation, some inhomogeneities can collapse into a
black-hole spacetime~\cite{garfinkle}. Furthermore, for the
inhomogeneities of the spacetime itself, energies in the form of
gravitational waves can also form a black-hole spacetime~\cite{nakao}.
As a consequence, at the onset of inflation, the distortion of the metric
by these inhomogeneities should be taken into account.
Furthermore, black holes or wormholes may be formed at peaks of
density fluctuations during inflation~\cite{nambu}.
Recenly,  in the context of inflation in string landscape,
it was shown that bubbles form by nucleation in metastable
vacua; our Universe may be one of these bubbles being separated
by domain walls from other bubbles and colliding with them~\cite{garriga}.

\subsection{Monopole or black hole}

\subsubsection{Schwarzschild-de Sitter spacetime}

Firstly, we consider the Schwarzschild-de Sitter (SdS) spacetime that
describes a black hole or a monopole sitting at the origin of an inflationary universe~\cite{SdS}.
In the static coordinate system, the line element of the SdS spacetime is given by
\begin{equation}
ds^{2}=-\left(1-{2GM\over r}-H^2r^2\right)dt^{2}+\left(1-{2GM\over
r}-H^2r^2\right)^{-1}dr^{2}+r^{2}d\Omega^{2}, \label{static}
\end{equation}
where $G=M_{\rm Pl}^{-2}$, $M$ is the mass of the black hole, and $H$ is the Hubble
parameter for inflation. Here we use the convention with
$c=\hbar=1$. As is well-known, the SdS metric has a black hole
horizon and a cosmological horizon.
In the static coordinates~(\ref{static}) an observer can only receive a signal
inside or just right on the cosmological horizon. This static
metric is insufficient for our purpose because in the cosmological
setting we aim at studying the temporal evolution of a Fourier
mode of the scalar quantum fluctuations that crosses the
cosmological horizon during inflation. Therefore, we will instead
use the planar coordinates for the SdS metric~\cite{shiromizu},
which is given by
\begin{equation}
ds^{2}=-f(r,\tau)d\tau^{2}+h(r,\tau) d{\vec x}^2,
\label{planar}
\end{equation}
where $r=|\vec x|$ and $d\tau=a^{-1}(\tau)dt$ is the conformal time defined in Eq.~(\ref{firstmetric}) with
$a(t)=e^{Ht}$. In Eq.~(\ref{planar}), for simplicity we have used the same notations,
$t$ and $r$, actually referring to different local coordinates
than those in Eq.~(\ref{static}).
The $f$ and $h$ functions are given by
\begin{equation}
f(r,\tau)=a^{2}(\tau)\left[1-\frac{GM}{2a(\tau)r}\right]^2
\left[1+\frac{GM}{2a(\tau)r}\right]^{-2},\quad
h(r,\tau)=a^{2}(\tau)\left[1+\frac{GM}{2a(\tau)r}\right]^{4},
\end{equation}
with the cosmic scale factor $a(\tau)=-1/(H\tau)$.
In these coordinates, the black hole horizon corresponds to
$r=GM/(2a)$ and the cosmological horizon is given by $r=a/H$.
For our purpose, we will restrict the range of validity of $\tau$ and $r$
to $-1/H<\tau<0$ and $GM/(2a)<r$.
Note that at late times (i.e., $\tau\rightarrow 0^{-}$) the planar coordinates
behave like a de Sitter expansion.

Let us define a dimensionless parameter $\epsilon=GMH$. Then, according to
Eq.~(\ref{pertmetric}), for small $\epsilon$ we have
\begin{equation}
g^{(1)}_{\mu\nu}=-\frac{2\epsilon}{H^{2}\tau r}{\rm diag}(1,1,1,1).
\end{equation}
We substitute $g^{(1)}_{\mu\nu}$ into Eq.~(\ref{source}) to get
\begin{equation}
J_{\vec k}(\tau,\vec{x})=-\frac{\epsilon k^{3/2} }{H \pi^{3/2}r}
e^{-ik\tau+i\vec{k}\cdot\vec{x}}\,\,\,.
\end{equation}
>From Eq.~(\ref{alpha}) and Eq.~(\ref{beta}) we obtain
\begin{eqnarray}
\alpha_{\vec{k},\vec{k}'}(\tau)&=&-\frac{i}{(2\pi)^{5/2}}\left(\frac{k}{k'}\right)^{3/2}
\int d^{3}x'\frac{e^{i(\vec{k}-\vec{k}')\cdot\vec{x}'}}{\mid\vec{x}'\mid}
\int_{\tau_{i}}^{\tau}d\tau'e^{-ik\tau'}(-k'\tau')^{3/2}H^{(2)}_{3/2}(-k'\tau')\nonumber\\
&=&-\frac{i}{\pi^{2}}\frac{1}{\mid\vec{k}-\vec{k}'\mid^{2}}\left(\frac{k}{k'}\right)^{3/2}
\left[\frac{-k+2k'+i(k-k')k'\tau'}{(k-k')^{2}}
e^{-i(k-k')\tau'}\right]_{\tau_{i}}^{\tau}\,\,\,,\label{alphaS}\\
\beta_{\vec{k},\vec{k}'}(\tau)&=&\frac{i}{\pi^{2}}\frac{1}{\mid\vec{k}+\vec{k'}\mid^{2}}
\left(\frac{k}{k'}\right)^{3/2}\left[\frac{k+2k'+i(k+k')k'\tau'}{(k+k')^{2}}
e^{-i(k+k')\tau'}\right]_{\tau_{i}}^{\tau}\,\,\,.\label{betaS}
\end{eqnarray}
Hence the first order correction of the two-point correlation function is given by
\begin{equation}
\Delta^{(1)}(\vec{x},\vec{x}')=\int d^{3}{k}
d^{3}{k}'e^{i\vec{k}\cdot\vec{x}+i\vec{k}'\cdot\vec{x}'}\Delta^{(1)}(\vec{k},\vec{k}')\,,
\end{equation}
where the power spectrum is
\begin{eqnarray}
\Delta^{(1)}(\vec{k},\vec{k}')&=& \frac{i\epsilon
H^{2}\tau^{2}}{16\pi^{5}\mid\vec{k}+\vec{k}'\mid^{2}}\frac{1}{(k-k')^{2}}\left(1+\frac{i}{k'\tau}-\frac{i}{k\tau}
+\frac{1}{kk'\tau^{2}}\right)\nonumber\\&
&\left\{-\frac{k^{2}}{k'^{2}}+\frac{k'^{2}}{k^{2}}+\frac{2k}{k'}-\frac{2k'}{k}
-i\frac{k}{k'}(k-k')\tau+i\frac{k'}{k}(k-k')\tau+\right.\nonumber\\& &
\left.e^{i(k-k')(\tau_{i}-\tau)}\left[-\frac{k'^{2}}{k^{2}}+\frac{k^{2}}{k'^{2}}+\frac{2k'}{k}-\frac{2k}{k'}
-i\frac{k'}{k}(k-k')\tau_{i}+i\frac{k}{k'}(k-k')\tau_{i}\right]\right\}\nonumber\\
& & +\frac{i\epsilon
H^{2}\tau^{2}}{16\pi^{5}\mid\vec{k}+\vec{k}'\mid^{2}}\frac{1}{(k+k')^{2}}
\left\{\left(1-\frac{i}{k'\tau}-\frac{i}{k\tau}-\frac{1}{kk'\tau^{2}}\right)\times \right.\nonumber\\
&&\left[-\frac{k^{2}}{k'^{2}}-\frac{2k}{k'}
+i(k+k')\frac{k}{k'}\tau+e^{i(k+k')(\tau_{i}-\tau)}
\left(\frac{k^{2}}{k'^{2}}+\frac{2k}{k'}-i(k+k')\frac{k}{k'}\tau_{i}\right)\right]\nonumber \\
&&+
\left(1+\frac{i}{k'\tau}+\frac{i}{k\tau}-\frac{1}{kk'\tau^{2}}\right)\times\left[\frac{k'^{2}}{k^{2}}+\frac{2k'}{k}
+i(k+k')\frac{k'}{k}\tau\right. \nonumber\\
&&\left.\left.-e^{-i(k+k')(\tau_{i}-\tau)}
\left(\frac{k'^{2}}{k^{2}}+\frac{2k'}{k}+i(k+k')\frac{k'}{k}\tau_{i}\right)\right]\right\}
\label{SdSHe}
\end{eqnarray}
When $\tau\rightarrow 0$, the power spectrum becomes
\begin{eqnarray}
\Delta^{(1)}(\vec{k},\vec{k}')&=&\frac{i\epsilon
H^{2}}{16\pi^{5}\mid\vec{k}+\vec{k}'\mid^{2}(k-k')^{2}k
k'}\times\nonumber\\
&&\left[e^{i(k-k')\tau_{i}}\left(-\frac{k'^{2}}{k^{2}}+\frac{k^{2}}{k'^{2}}+\frac{2k'}{k}-\frac{2k}{k'}
-i\frac{k'}{k}(k-k')\tau_{i}+i\frac{k}{k'}(k-k')\tau_{i}\right)\right]\nonumber\\
&&-\frac{i\epsilon
H^{2}}{16\pi^{5}\mid\vec{k}+\vec{k}'\mid^{2}(k+k')^{2}kk'}\left\{\left[e^{i(k+k')\tau_{i}}
\left(\frac{k^{2}}{k'^{2}}+\frac{2k}{k'}-i(k+k')\frac{k}{k'}\tau_{i}\right)\right]\right.\nonumber\\
&&\,\,\,\,\,\,\,\,\,\,\,\,\,\,\,\,\,\,\,\,\,\,\,\,\left. -\left[e^{-i(k+k')\tau_{i}}\left(\frac{k'^{2}}{k^{2}}+\frac{2k'}{k}+i(k+k')\frac{k'}{k}\tau_{i}\right)\right]\right\}
\end{eqnarray}
This spectrum still looks complicated. Let us examine a special case with
$\mid\vec{k}'\mid=\mid\vec{k}\mid$ at $\tau=0$. Then the power
spectrum is given by
\begin{eqnarray}
\Delta^{(1)}(\vec{k},\vec{k}')=
\Delta^{(1)}(k,\theta)&=&\frac{\epsilon H^{2}}{64\pi^{5}}\frac{1}{k^{6}(1+\cos\theta)}
\left[3\sin(2k\tau_{i})-2k\tau_{i}\cos(2k\tau_{i})-4k\tau_{i}\right]\,\,\,,
\end{eqnarray}
where $k=\mid\vec{k}\mid$ and $\theta$ is the separation angle between $\vec{k}$ and $\vec{k}$.

In Eq.~(\ref{SdSHe}), if we take the limit, $\tau_{i}=-\infty(1+i0^+)$, we would obtain the power spectrum as
\begin{equation}
\Delta^{(1)}(\vec{k},\vec{k}')=\frac{\epsilon H^2}{8\pi^5}\frac{\tau^3}{(k+k')\mid\vec{k}+\vec{k}'\mid^{2}},
\label{bhpert}
\end{equation}
which has a simple form but it vanishes as $\tau\rightarrow 0$.

\subsubsection{Reissner-Nordstr\"{o}m-de Sitter spacetime}

The other interesting example is to consider many black holes in an
expanding universe that should be more likely to occur in the early Universe. We
will calculate the power spectrum of scalar field fluctuations in the presence of many black holes.
The Majumdar-Papapetrou solution describes a collection of charged black
holes in Minkowskian space whose gravitational attractions are balanced by
electrostatic repulsions~\cite{MP}. Kastor and Trashen~\cite{RNdS} extended
the Majumdar-Papapetrou solution to a spacetime with a cosmological
constant which is called Reissner-Nordstr\"{o}m-de Sitter (RNdS) solution.
This solution can be rewritten in a simple form in the cosmological or planar
coordinates~\cite{RNdS}.  Our results of the power spectrum for the
SdS black hole can be straightforwardly applied to
this case. In the cosmological coordinates, the
solution of the Einstein equation for a distribution of RNdS extremal black holes  is
written in the following form~\cite{RNdS},
\begin{equation}
ds^{2}=-a^{2}\Omega^{-2} d \tau^{2}+a^{2}\Omega^{2}d\vec{x}^{2},
\end{equation}
where $\Omega=1+\sum_{i}Gm_{i}/(a r_{i})$, $m_i$ is the $i$th black hole mass, and
$r_i=\mid \vec{x} - \vec{x}_i\mid$. Here we have omitted the gauge field solution
of the Maxwell equation since it is irrelevant to the present consideration.

Here we assume $m_i=m$ and consider a small parameter $\epsilon=GHm$.
As such, similar to the SdS black hole case, we have
\begin{equation}
g^{(1)}_{\mu\nu}=-\sum_i\frac{2\epsilon}{H^{2}\tau r_i}{\rm diag}(1,1,1,1),
\end{equation}
and the source term is then given by
\begin{equation}
J_{\vec k}(\tau,\vec{x})=-\sum_i\frac{\epsilon k^{3/2} }{H \pi^{3/2}r_i}
e^{-ik\tau+i\vec{k}\cdot\vec{x}}\,\,\,.
\end{equation}
For a continuous density function of black holes, denoted by $\rho(\vec{x})$,
the source term can be written as
\begin{equation}
J_{\vec k}(\tau,\vec{x})=-\frac{\epsilon k^{3/2} }{H \pi^{3/2}}
e^{-ik\tau+i\vec{k}\cdot\vec{x}}\int d^{3}x' \frac{\rho(\vec {x}')}{\mid\vec{x}-\vec{x}'\mid}\,\,\,.
\end{equation}
Following the same steps as in the SdS black hole, we obtain for the RNdS black holes
\begin{equation}
\alpha^{RN}_{\vec{k},\vec{k}'}(\tau)=-\frac{i}{(2\pi)^{5/2}}\left(\frac{k}{k'}\right)^{3/2}
\int_{\tau_{i}}^{\tau}d\tau'e^{-ik\tau'}(-k'\tau')^{3/2}H^{(2)}_{3/2}(-k'\tau')
\int d^{3}x d^{3}x'\frac{\rho(\vec{x}')e^{i(\vec{k}-\vec{k}')\cdot\vec{x}}}
{\mid\vec{x}-\vec{x}'\mid}.
\end{equation}
Let $\vec{K}=\vec{k}-\vec{k}'$ and $\vec{y}=\vec{x}-\vec{x}'$. Then, after doing the $\vec{y}$-integration,
we obtain
\begin{equation}
\alpha^{RN}_{\vec{k},\vec{k}'}(\tau)=F(\vec{K}) \alpha^{S}_{\vec{k},\vec{k}'}(\tau),\quad
F(\vec{K})=\int d^{3}x\rho(\vec{x})e^{i\vec{K}\cdot\vec{x}},
\end{equation}
where $\alpha^{S}_{\vec{k},\vec{k}'}$ is given by Eq.~(\ref{alphaS}) and
$F(\vec{K})$ is the so-called form factor, which is the Fourier transform of
the black-hole density function. Similarly, we have $\beta^{RN}_{\vec{k},\vec{k}'}
=F(\vec{K}') \beta^{S}_{\vec{k},\vec{k}'}$, where
$\vec{K}'=\vec{k}+\vec{k}'$ and $\beta^{S}_{\vec{k},\vec{k}'}$ is given by Eq.~(\ref{betaS}).
And $\Delta^{(1)}_{RN}(\vec{k},\vec{k}')=F^{\ast}(\vec{K}')\Delta^{(1)}_S(\vec{k},\vec{k}')$, where $\Delta^{(1)}_S(\vec{k},\vec{k}')$ is given by Eq.~(\ref{SdSHe}).

Let us work out an example of the form factor. Consider a lattice distribution of black holes that has
\begin{equation}
\rho(\vec{x})=\sum_{lmn}\delta(\vec{x}-\vec{x}_{lmn})\,, \quad
\vec{x}_{lmn}=la\hat{x}+mb\hat{y}+nc\hat{z}\,\,,
\end{equation}
where $a$, $b$, $c$ are lattice sizes, and $\emph{l}$, $\emph{m}$, $\emph{n}$ are integers,
ranging from $-N$ to $N$.
Then, the form factor of this distribution is
\begin{equation}
F(\vec{K})=\sum_{lmn}e^{i\vec{K}\cdot\vec{x}_{lmn}}
=\sum_{l}e^{i laK_{x}}\sum_{m}e^{i mbK_{y}}\sum_{n}e^{i nbK_{z}},
\end{equation}
where
\begin{equation}
\sum_{l}e^{i laK_{x}}=\frac{2\cos(\frac{N}{2}aK_{x})\sin(\frac{N+1}{2}aK_{x})}{\sin(\frac{1}{2}{aK_{x})}}-1,
\end{equation}
and similar expressions for the remaining summations over $m$ and $n$.

\subsection{Cosmic string}

We consider a cosmic string in an inflating universe. This
cosmic string may be formed as the Universe went through a phase
transition. The metric for an inflationary spacetime with an infinitely long
string passing through the origin along the $z$-axis is depicted as~\cite{String}
\begin{equation}
ds^{2}=-dt^{2}+a(t)^{2}[d\rho^{2}+\rho^{2}(1-4G\mu)^{2}d\phi^{2}+dz^{2}]\,\,\,.
\label{stringmetric}
\end{equation}
It can be decomposed into diagonal terms $g^{(0)}_{\mu\nu}$ and off-diagonal terms $g^{(1)}_{\mu\nu}$:
\begin{equation}
ds^{2}=a^{2}(-d\tau^{2}+dx^{2}+dy^{2}+dz^{2})-\frac{2\epsilon
a^{2}}{\rho^{2}}(ydx-xdy)^{2}\,\,\,,
\label{stringxyz}
\end{equation}
where $\epsilon=4G\mu$.
Substituting $g^{(0)}_{\mu\nu}$ and $g^{(1)}_{\mu\nu}$ into
Eq.~(\ref{source}), we have the source term for the cosmic-string spacetime,
\begin{equation}
J_{\vec{k}}(\tau,\vec{x})=2\epsilon
a^{2}(\tau)\frac{(xk_{y}-yk_{x})^{2}+i(xk_{x}+yk_{y})}{x^{2}+y^{2}}\varphi_{k}(\tau)e^{i\vec{k}\cdot\vec{x}}\,\,\,.
\end{equation}
Then we plug this into Eq.~(\ref{alpha}) to obtain
\begin{eqnarray}
\alpha_{\vec{k},\vec{k}'}(\tau)&=&\frac{\delta(k_{z}-k'_{z})}{(2\pi)^{2}k^{3/2}k'^{3/2}}
\left[2\pi^{2}\delta(\vec{k}_{\perp}-\vec{k}'_{\perp})\vec{k}^{2}_{\perp}
-\frac{2\pi\vec{k}_{\perp}\cdot\vec{k}'_{\perp}}{(\vec{k}_{\perp}-\vec{k}'_{\perp})^{2}}
+4\pi\frac{[(\vec{k}_{\perp}-\vec{k}'_{\perp})\cdot\vec{k}_{\perp}]^{2}}{(\vec{k}_{\perp}-\vec{k}'_{\perp})^{4}}\right]\nonumber\\
&&\times\left[\left(\frac{k k'}{k-k'}+\frac{i}{\tau}\right)e^{-i(k-k')\tau}-\left(\frac{k
k'}{k-k'}+\frac{i}{\tau_{i}}\right)e^{-i(k-k')\tau_{i}}\right]\,\,\,,\nonumber\\
\beta_{\vec{k},\vec{k}'}(\tau)&=&\frac{\delta(k_{z}+k'_{z})}{(2\pi)^{2}k^{3/2}k'^{3/2}}
\left[2\pi^{2}\delta(\vec{k}_{\perp}+\vec{k}'_{\perp})\vec{k}^{2}_{\perp}
+\frac{2\pi\vec{k}_{\perp}\cdot\vec{k}'_{\perp}}{(\vec{k}_{\perp}+\vec{k}'_{\perp})^{2}}
+4\pi\frac{[(\vec{k}_{\perp}+\vec{k}'_{\perp})\cdot\vec{k}_{\perp}]^{2}}{(\vec{k}_{\perp}+\vec{k}'_{\perp})^{4}}\right]\nonumber\\
&&\times\left[\left(-\frac{kk'}{k+k'}+\frac{i}{\tau}\right)e^{-i(k+k')\tau}-\left(-\frac{k
k'}{k+k'}+\frac{i}{\tau_{i}}\right)e^{-i(k+k')\tau_{i}}\right]\,\,\,.
\end{eqnarray}
Now we are ready to calculate $\Delta^{(1)}(\vec{k},\vec{k}')$. By
substituting $\alpha_{\vec{k},\vec{k}'}(\tau)$ and
$\beta_{\vec{k},\vec{k}'}(\tau)$ into Eq.~(\ref{kspectrum}) and
letting $\tau_{i}=-\infty(1+i0^+)$, we obtain
\begin{eqnarray}
\Delta^{(1)}(\vec{k},\vec{k}')& = &\frac{\epsilon
H^{2}}{(2\pi)^{6}}\frac{1+k^{2}\tau^{2}}{k^{3}(k^2-k'^2)}2\pi\delta(k_{z}+k'_{z})\times\nonumber\\
& &
\left\{2\pi^{2}\delta(\vec{k}_{\perp}+\vec{k}'_{\perp})\vec{k}^{2}_{\perp}
+2\pi\frac{\vec{k}_{\perp}\cdot\vec{k}'_{\perp}}{(\vec{k}_{\perp}+\vec{k}'_{\perp})^{2}}
+\frac{4\pi[(\vec{k}_{\perp}+\vec{k}'_{\perp})\cdot\vec{k}_{\perp}]^{2}}{(\vec{k}_{\perp}+\vec{k}'_{\perp})^{4}}\right\}\nonumber\\
&&\ \ \ \ \ +(\vec{k}\leftrightarrow\vec{k}')
\label{skspectrum}\,\,\,.
\end{eqnarray}
The delta function $\delta(k_{z}+k'_{z})$ implies that the
translation along the $z$-axis is unbroken as it is expected.
Furthermore, the first term in the large brackets of
Eq.~(\ref{skspectrum}) preserves translational invariance while the
second and third terms are not. This feature has also been pointed
out by Ref.~\cite{TsengWise,string}. Eq.~(\ref{skspectrum}) can be
further symmetrized as
\begin{eqnarray}
\Delta^{(1)}(\vec{k},\vec{k}')&=&  \frac{\epsilon
H^{2}\delta(k_{z}+k'_{z})}{(2\pi)^{5}}\cdot\frac{k^{2}+k'^{2}+kk'+k^{2}k'^{2}\tau^{2}}{(k+k')k^{3}k'^{3}}\times\nonumber\\
& &
\left[(2\pi)^{2}\delta(\vec{k}_{\perp}+\vec{k}'_{\perp})\frac{\vec{k}_{\perp}\cdot\vec{k}'_{\perp}}{2}
-2\pi\frac{\vec{k}_{\perp}\cdot\vec{k}'_{\perp}}{(\vec{k}_{\perp}+\vec{k}'_{\perp})^{2}}
+\frac{4\pi(\vec{k}_{\perp}+\vec{k}'_{\perp})\cdot\vec{k}_{\perp}(\vec{k}_{\perp}+\vec{k}'_{\perp})\cdot\vec{k}'_{\perp}}
{(\vec{k}_{\perp}+\vec{k}'_{\perp})^{4}}\right]\nonumber\\
&&+  \frac{\epsilon
H^{2}\delta(k_{z}+k'_{z})}{(2\pi)^{5}}\cdot\frac{2\pi}{(k+k')k^{3}k'^{3}}\times\nonumber\\
& &
\left[(k^{2}+k'^{2}-kk'+k^{2}k'^{2}\tau^{2})
\frac{(k+k')^2}{(\vec{k}_{\perp}+\vec{k}'_{\perp})^{2}}
-(k^{2}+k'^{2}+kk'+k^{2}k'^{2}\tau^{2})\right]
\,\,\,.
\label{skspectrum2}
\end{eqnarray}

\subsection{Domain wall}

Consider the metric~\cite{Dwall}
which describes an infinitely large flat domain wall in an
inflationary spacetime. The metric for such case with the wall placed at $z=0$ is
\begin{equation}
ds^{2}=\frac{1}{\alpha^{2}(\tau+\epsilon\mid z
\mid)^{2}}(-d\tau^{2}+dx^{2}+dy^{2}+dz^{2})\,\,\,,
\end{equation}
where $1/\alpha^{2}=(1-\epsilon^{2})/H^{2}$ and $-1<\epsilon\leq0$.

To facilitate the perturbative approach that we are employing so far, we make the following coordinate transformation, $\tau\rightarrow\tilde{\tau}=\tau+\epsilon|z|$. Then the metric becomes
\begin{equation}
ds^{2}=\frac{1-\epsilon^{2}}{H^{2}\tau^{2}}(-d\tau^{2}\pm 2\epsilon d\tau dz+dx^{2}+dy^{2}+(1-\epsilon^{2})dz^2),\label{dwall}
\end{equation}
where $+$ and $-$ signify the regions with $z>0$ and $z<0$, respectively. Note that for simplicity we have dropped the $\ \tilde{}\ $ in $\tau$. Expanding the metric to first order in $\epsilon$, we have
\begin{equation}
g^{(1)}_{\tau z}=g^{(1)}_{z\tau}=\pm\frac{\epsilon}{H^{2}\tau^{2}},
\end{equation}
as the only non-vanishing components.
Plugging $g^{(1)}_{\mu\nu}$ into Eq.~(\ref{source}), we obtain the source term
$J_{\vec{k}}$ for both $z>0$ and $z<0$ regions as
\begin{equation}
J_{\vec{k}}(\tau,\vec{x})=\mp\frac{
\epsilon k_{z}}{2H\pi^{3/2}k^{3/2}\tau^{3}}(1+ik\tau-k^{2}\tau^{2})e^{-ik\tau+i\vec{k}\cdot\vec{x}}
\end{equation}
Then, according to Eq.~(\ref{alpha}) and Eq.~(\ref{beta}), we can
write down $\alpha_{\vec{k},\vec{k}'}(\tau)$ and
$\beta_{\vec{k},\vec{k}'}(\tau)$ as
\begin{eqnarray}
\alpha_{\vec{k},\vec{k}}(\tau)&=&-\frac{ik_{z}}{\pi(k_{z}-k'_{z})k^{3/2}k'^{3/2}}
\delta(\vec{k}_{\perp}-\vec{k}'_{\perp})\times\nonumber\\
&&\ \ \ \ \ \left[\left(\frac{1}{2\tau'^{2}}+\frac{i(k-k')}{2\tau'}
+\frac{k^{2}k'}{k-k'}\right)e^{-i(k-k')\tau'}+\frac{1}{2}(k^{2}-k'^{2}){\rm Ei}[-i(k-k')\tau']\right]_{\tau_{i}}^{\tau}\nonumber\\
\beta_{\vec{k},\vec{k}}(\tau)&=&-\frac{ik_{z}}{\pi(k_{z}+k'_{z})k^{3/2}k'^{3/2}}
\delta(\vec{k}_{\perp}+\vec{k}'_{\perp})\times\nonumber\\
&&\ \ \ \ \ \left[\left(\frac{1}{2\tau'^{2}}+\frac{i(k+k')}{2\tau'}
-\frac{k^{2}k'}{k+k'}\right)e^{-i(k+k')\tau'}+\frac{1}{2}(k^{2}-k'^{2}){\rm Ei}[-i(k+k')\tau']\right]_{\tau_{i}}^{\tau}\nonumber\\
\end{eqnarray}
Hence, substituting $\alpha_{\vec{k},\vec{k}'}(\tau)$ and
$\beta_{\vec{k},\vec{k}'}(\tau)$ into Eq.~(\ref{kspectrum}), we can construct the first order correction
$\Delta^{(1)}(\vec{k},\vec{k}')$. Taking $\tau_{i}\rightarrow-\infty$ and neglecting wildly oscillatory terms, we have
\begin{eqnarray}
\Delta^{(1)}(\vec{k},\vec{k}')
&=&-\frac{\epsilon H^{2}\tau(kk_{z}+k'k'_{z})\delta(\vec{k}_{\perp}+\vec{k}'_{\perp})}{16\pi^{4}k^{2}k'^{2}(k_{z}+k'_{z})}\nonumber\\
&&+\frac{i\epsilon H^{2}\tau^{2}(k_{z}-k'_{z})\delta(\vec{k}_{\perp}+\vec{k}'_{\perp})}
{64\pi^{4}k^{2}k'^{2}}\times\nonumber\\
&&\ \ \ \ \ \ \ \left[(k_{z}+k'_{z})\left(1-\frac{i}{k\tau}+\frac{i}{k'\tau}+\frac{1}{kk'\tau^{2}}\right){\rm Ei}[i(k-k')\tau]e^{-i(k-k')\tau}\right.\nonumber\\
&&\ \ \ \ \ \ \ -(k_{z}+k'_{z})\left(1+\frac{i}{k\tau}-\frac{i}{k'\tau}+\frac{1}{kk'\tau^{2}}\right){\rm Ei}[-i(k-k')\tau]e^{i(k-k')\tau}\nonumber\\
&&\ \ \ \ \ \ \ +(k_{z}-k'_{z})\left(1-\frac{i}{k\tau}-\frac{i}{k'\tau}-\frac{1}{kk'\tau^{2}}\right){\rm Ei}[i(k+k')\tau]e^{-i(k+k')\tau}\nonumber\\
&&\ \ \ \ \ \ \ \left.-(k_{z}-k'_{z})\left(1+\frac{i}{k\tau}+\frac{i}{k'\tau}-\frac{1}{kk'\tau^{2}}\right){\rm Ei}[-i(k+k')\tau]e^{i(k+k')\tau}\right],\label{wallpert}
\end{eqnarray}
where ${\rm Ei}(z)=-\int_{-z}^{\infty}dt\,(e^{-t}/t)$ is the exponential integral function. To obtain the above expression we have symmetrized with respect to $\vec{k}$ and $\vec{k}'$ since $\Delta(\vec{x},\vec{x}')$ in Eq.~(\ref{twopointcorrelator}) should be symmetric with respect to $\vec{x}$ and $\vec{x}'$.
Again the presence of the delta function $\delta(\vec{k}_{\perp}+\vec{k}'_{\perp})$ indicates that there is translational invariance in the $x$-$y$ plane, while the invariance along the $z$-axis is broken.

When we take $\tau\rightarrow 0$, we have, for $k_{z}>k_{z}'$,
\begin{eqnarray}
\Delta^{(1)}(\vec{k},\vec{k}')
&=&-\frac{\epsilon H^{2}k_{z}'(k^{2}-k'^{2})\delta(\vec{k}_{\perp}+\vec{k}'_{\perp})}{32\pi^{3}k^{3}k'^{3}(k_{z}+k_{z}')},
\end{eqnarray}
while for $k_{z}<k_{z}'$,
\begin{eqnarray}
\Delta^{(1)}(\vec{k},\vec{k}')
&=&-\frac{\epsilon H^{2}k_{z}(k^{2}-k'^{2})\delta(\vec{k}_{\perp}+\vec{k}'_{\perp})}{32\pi^{3}k^{3}k'^{3}(k_{z}+k_{z}')}.
\end{eqnarray}
Note that for $k_{z}=k_{z}'$ or $k=k'$, $\Delta^{(1)}(\vec{k},\vec{k}')=0$.

\section{Conclusions}
\label{conclusion}

We have presented a perturbation method to compute the effects of the presence of cosmic defects in the de Sitter space to the quantum fluctuations of a free massless scalar field.
The method is valid as long as the metric distortion from the cosmic defects to the de Sitter space is small.
In particular, we have computed the first order contribution in the two-point correlation function of the scalar field.
The calculation can be easily generalized to a vector field or a gravitational wave.
Our work is a realization of the general discussions in Ref.~\cite{carroll}
about the potentially observable effects of a small violation of
translational invariance during inflation, as characterized by the presence of a preferred point, line, or plane.
It would be very interesting to study the implications of our results to the metric perturbation in inflation and their
signatures on the CMB anisotropies.

One thing should be noted is that, as shown in the Appendix, the
in-in formalism and the perturbative method used in this paper should be equivalent
although they give slightly different results in certain cases. In the usual in-in formalism, the Hamitonian
of the considered quantum system is assumed to be finite. This assumption is invalid in the present work
where the string or the domain wall is of infinite extent and thus
the Hamitonian over the whole space is undefined.
However, in the black hole case the interaction term has $1/r$ dependence.
This renders the Hamitonian finite and indeed we have found that our perturbative result
agrees with that in the in-in calculation. Although the explicit forms of the power spectra in the string and the domain
wall cases are slightly different in both methods, the two-point functions that require integrating over momenta seem
to be indistinguishable. More work needs to be done to understand these issues better.

\section*{Acknowledgments}

We would like to thank the authors in ~\cite{solomonetal} for their
correspondence about the in-in formalism of the domain wall case.
This work was supported in part by the National Science Council,
Taiwan, ROC under the Grants No. NSC101-2112-M-001-010-MY3 (K.W.N.)
, NSC102-2112-M-032-002-MY3 (H.T.C.) and NSC100-2112-M-032-001-MY3
(I.C.W.). HTC is also supported in part by the National Center for
Theoretical Sciences (NCTS). HTC would like to thank the hospitality
of the Theory Group of the Institute of Physics at the Academia
Sinica, Republic of China, where part of this work was done.

\section*{APPENDIX A: In-in formalism}

In this Appendix, we will use the in-in formalism (see Ref.~\cite{weinberg}) to derive the power spectra.
In the perturbative method that we have used above, the results for the power spectra depend on the initial time
when the source is turned on. This is necessary for the case in which we are concerned about the effect of cosmic defects
on the modes of inflaton fluctuations that leave the horizon just after the onset of inflation. As we have mentioned above, our results
may have interesting implications to cosmological observations on large angular scales in
inflation models with finite duration lasting for about 60 e-foldings. On the other hand, in the in-in formalism the initial
time is usually set at the infinite past in such a way that the early time history is exponentially suppressed; consequently, the system
reaches a steady state and thus the power spectra do not depend on the initial condition. Below we will use the same initial time setting to derive the power spectra and compare them with those obtained in our perturbative method.

\subsection{Black hole}

Here we compute the power spectrum $\Delta^{(1)}(\vec{k},\vec{k}')$ in the in-in formalism,
using the Hamiltonian interaction picture. Using the planar coordinates for the
SdS metric in Eq.~(\ref{planar}), the Lagrangian density for a massless inflaton is given by
\begin{eqnarray}
\mathcal{L}& =
&-\frac{\sqrt{-g}}{2}g^{\mu\nu}\partial_{\mu}\phi\partial_{\nu}\phi\nonumber\\
& =
&\frac{a^{2}}{2}\frac{\left(1+\frac{GM}{2ar}\right)^{7}}{\left(1-\frac{GM}{2ar}\right)}(\partial_{\tau}\phi)^{2}-
\frac{a^{2}}{2}\left(1-\frac{GM}{2ar}\right)\left(1+\frac{GM}{2ar}\right)
(\vec{\nabla}\phi)^{2}.
\end{eqnarray}
Let us define the conjugate momentum,
\begin{equation}
\Pi= \frac{\partial\mathcal{L}}{\partial\partial_{\tau}\phi}=
\frac{\left(1+\frac{GM}{2ar}\right)^{7}}{\left(1-\frac{GM}{2ar}\right)}a^2\partial_{\tau}\phi\,.
\end{equation}
Then, the Hamiltonian would be given by
\begin{eqnarray}
\emph{H}& = & \int d^{3}x
(\Pi\partial_{\tau}\phi-\mathcal{L})\nonumber\\
& = &\int d^{3}x
\frac{1}{2a^2}\frac{\left(1-\frac{GM}{2ar}\right)}{\left(1+\frac{GM}{2ar}\right)^{7}}\Pi^{2}+
\frac{a^{2}}{2}\left(1-\frac{GM}{2ar}\right)\left(1+\frac{GM}{2ar}\right)
(\vec{\nabla}\phi)^{2}.\nonumber\\
\end{eqnarray}
Expanding $\emph{H}$ in terms of ${GM}/({2ar})$, to the first order we have
\begin{eqnarray}
\emph{H}_{0}&=&\int
d^{3}x \frac{a^{2}}{2}\left[\frac{\Pi^2}{a^4}+(\vec{\nabla}\phi)^{2}\right]\,,
\\ \emph{H}_{I}& = -&\int
d^{3}x \frac{a^{2}}{2}\left(\frac{4GM}{ar}\frac{\Pi^2}{a^4}\right)\,. \label{HI}
\end{eqnarray}
Now we calculate the two-point correlation function following the in-in formalism~\cite{weinberg}:
\begin{eqnarray}
\langle\phi(\vec{x},\tau)\phi(\vec{y},\tau)\rangle & \simeq &
\langle\phi_{I}(\vec{x},\tau)\phi_{I}(\vec{y},\tau)\rangle\nonumber\\
 & & +i\int_{-\infty}^{\tau}d\tau' e^{-\epsilon'\mid \tau'\mid}
 \langle\left[\emph{H}_{I}(\tau'),\phi_{I}(\vec{x},\tau)\phi_{I}(\vec{y},\tau)\right]\rangle,
 \label{inexpand}
\end{eqnarray}
where $\epsilon'$ is an infinitesimal positive parameter that suppresses the infinite past
of the time integration. The field  $\phi_{I}$ is the interaction-picture one governed by the
unperturbed Hamiltonian $\emph{H}_{0}$, obeying the free scalar field equation of motion,
\begin{equation}
\left(\partial^{2}_{\tau}-\frac{2}{\tau}\partial_{\tau}-\vec{\nabla}^{2}\right)\phi_I=0\,,
\end{equation}
which has a plane-wave solution,
\begin{eqnarray}
\phi_{I}& =
&\int\frac{d^{3}k}{(2\pi)^{3}}e^{i\vec{k}\cdot\vec{x}}
\left[\phi_{k}(\tau)a_{\vec{k}}+\phi^{\ast}_{k}(\tau)a^{\dagger}_{-\vec{k}}\right],\nonumber\\
\phi_{k}(\tau)&=& -\frac{H}{\sqrt{2k}}e^{-ik\tau}\left(\tau-\frac{i}{k}\right),\nonumber\\
&&[a_{\vec{k}}, a^{\dagger}_{\vec{q}}]=(2\pi)^3\delta(\vec{k}-\vec{q}).
\label{freefield}
\end{eqnarray}
Inserting this free field solution into Eq.~(\ref{HI}) where $\Pi=a^2\partial_\tau\phi_I$, the interaction-picture Hamitonian $H_{I}(\tau)$ in Eq.~(\ref{inexpand}) is given by
\begin{eqnarray}
\emph{H}_{I}(\tau)&=&\frac{2GM}{H\tau}\int
d^{3}x{1\over r}\int\frac{d^{3}k}{(2\pi)^{3}}e^{i\vec{k}\cdot\vec{x}}
\int\frac{d^{3}q}{(2\pi)^{3}}e^{i\vec{q}\cdot\vec{x}}\nonumber\\
&
&\times \left[\phi'_{k}(\tau)a_{\vec{k}}+\phi'^{\ast}_{k}(\tau)a^{\dag}_{-\vec{k}}\right]
\left[\phi'_{q}(\tau)a_{\vec{q}}+\phi'^{\ast}_{q}(\tau)a^{\dag}_{-\vec{q}}\right],
\end{eqnarray}
where the prime denotes the differentiation with respect to $\tau$.
Then, we can calculate the commutator,
\begin{eqnarray}
&
&\langle\left[\emph{H}_{I}(\tau'),\phi_{I}(\vec{x},\tau)\phi_{I}(\vec{y},\tau)\right]\rangle\nonumber\\
 & =&\frac{4GM}{H\tau'}\int\frac{d^{3}k}{(2\pi)^{3}} \int\frac{d^{3}q}{(2\pi)^{3}}
\frac{4\pi}{\mid\vec{k}+\vec{q}\mid^2} e^{-i\vec{k}\cdot\vec{x}-i\vec{q}\cdot\vec{y}}\times\nonumber\\
&
&\left\{\left[\frac{-ikH}{\sqrt{2k}}e^{-ik\tau'}\left(\tau'-\frac{i}{k}\right)+\frac{H}{\sqrt{2k}}e^{-ik\tau'}\right]
\left[\frac{-iqH}{\sqrt{2q}}e^{-iq\tau'}\left(\tau'-\frac{i}{q}\right)+\frac{H}{\sqrt{2q}}e^{-iq\tau'}\right]\right.\nonumber\\
& &
\times\left.\frac{H}{\sqrt{2k}}e^{ik\tau}\left(\tau+\frac{i}{k}\right)\frac{H}{\sqrt{2q}}e^{iq\tau}\left(\tau+\frac{i}{q}\right)
-h.c.\right\}.
\end{eqnarray}
Now we substitute the commutator into the second term of
Eq.~(\ref{inexpand}) which is the first order correction and perform
the integration over $\tau'$. Thus, we obtain
\begin{eqnarray}
\Delta^{(1)}\langle\phi(\vec{x},\tau)\phi(\vec{y},\tau)\rangle& = &
\langle\phi(\vec{x},\tau)\phi(\vec{y},\tau)\rangle - \langle\phi_{I}(\vec{x},\tau)\phi_{I}(\vec{y},\tau)\rangle\nonumber\\
 & = &8\pi\epsilon H^2
 \int\frac{d^{3}k}{(2\pi)^{3}}\int\frac{d^{3}q}{(2\pi)^{3}}e^{-i\vec{k}\cdot\vec{x}-i\vec{q}\cdot\vec{y}}
 \frac{\tau^3}{(k+q)\mid\vec{k}+\vec{q}\mid^{2}}\,,
 \end{eqnarray}
 where $\epsilon=GMH$ and the power spectrum is
 \begin{equation}
\Delta^{(1)}(\vec{k},\vec{q})=\frac{\epsilon H^2}{8\pi^5}\frac{\tau^3}{(k+q)\mid\vec{k}+\vec{q}\mid^{2}}\,\,,
 \end{equation}
which is exactly the same as the perturbative result obtained in Eq.~(\ref{bhpert}).

\subsection{Cosmic string}

Using the metric~(\ref{stringxyz}) the Lagrangian density of a massless scalar field is given by
\begin{equation}
{\cal L}=-\frac{a^2}{2}(1-2\epsilon)^{1\over2}\left[-(\partial_{\tau}\phi)^{2}+(\vec{\nabla}\phi)^{2}+\left(\frac{2\epsilon}{1-2\epsilon}\right)
\left(\frac{1}{x^2+y^2}\right)(y\partial_{x}\phi-x\partial_{y}\phi)^{2}\right].
\end{equation}
Then the conjugate momentum is
\begin{equation}
\Pi= a^2(1-2\epsilon)^{1\over2}\partial_{\tau}\phi\,,
\end{equation}
and the Hamiltonian is
\begin{eqnarray}
H&=&\int d^{3}x\frac{a^2}{2}(1-2\epsilon)^{1\over2}\left[\frac{\Pi^2}{a^4(1-2\epsilon)}+(\vec{\nabla}\phi)^{2}+\left(\frac{2\epsilon}{1-2\epsilon}\right)
\left(\frac{1}{x^2+y^2}\right)(y\partial_{x}\phi-x\partial_{y}\phi)^{2}\right]\nonumber\\
&=&H_0+H_I.
\end{eqnarray}
To the linear order in $\epsilon$, we have the free Hamitonian and the interaction Hamitonian respectively,
\begin{eqnarray}
H_0&=&\int d^{3}x\frac{a^2}{2}\left[\frac{\Pi^2}{a^4}+(\vec{\nabla}\phi)^{2}\right],\\
H_I&=&\epsilon\int d^{3}x \frac{a^2}{2}\left[\frac{\Pi^2}{a^4}-(\vec{\nabla}\phi)^{2}+\frac{2}{x^2+y^2}(y\partial_{x}\phi-x\partial_{y}\phi)^{2}\right] .\label{HIstring}
\end{eqnarray}
Inserting the free field solution~(\ref{freefield}) into Eq.~(\ref{HIstring}) where $\Pi=a^2\partial_\tau\phi_I$, the commutator is given by
\begin{eqnarray}
& &\langle[\emph{H}_{I}(\tau'),\phi_{I}(\vec{x},\tau)\phi_{I}(\vec{y},\tau)]\rangle\nonumber\\
&=&i\epsilon H^{2}\int d^{3}x'\int\frac{d^{3}k}{(2\pi)^{3}}
\int\frac{d^{3}q}{(2\pi)^{3}}e^{i\vec{k}\cdot(\vec{x'}-\vec{x})+i\vec{q}\cdot(\vec{x'}-\vec{y})}\times\nonumber\\
&&\left\{-\frac{1}{2}\left[(\tau^{2}-\frac{1}{kq})\sin((k+q)(\tau-\tau'))+\frac{\tau(k+q)}{kq}\cos((k+q)(\tau-\tau'))\right]\right.\nonumber\\
&&+\frac{(k_{x}q_{x}+k_{y}q_{y}+k_{z}q_{z})}{2\tau'^{2}kq}\left[((\tau^{2}-\frac{1}{kq})(\tau'^{2}-\frac{1}{kq})+\tau\tau'\frac{(k+q)^{2}}{k^{2}q^{2}})\sin((k+q)(\tau-\tau'))\right.\nonumber\\
&&+\left.\frac{k+q}{kq}(\tau(\tau'^{2}-\frac{1}{kq})-\tau'(\tau^{2}-\frac{1}{kq}))\cos((k+q)(\tau-\tau'))\right]\nonumber\\
&&-\frac{(y'k_{x}-x'k_{y})(y'q_{x}-x'q_{y})}{\tau'^{2}kq(x'^{2}+y'^{2})}
\cdot\left[((\tau'^{2}-\frac{1}{kq})(\tau^{2}-\frac{1}{kq})+\tau\tau'(\frac{k+q}{kq})^{2})\sin((k+q)(\tau-\tau'))\right.\nonumber\\
&&+\left.\frac{k+q}{kq}(\tau(\tau'^{2}-\frac{1}{kq})-\tau'(\tau^{2}-\frac{1}{kq}))\cos((k+q)(\tau-\tau'))\right]
\end{eqnarray}

The calculation for a cosmic string in the metric
Eq.~(\ref{stringmetric}) was finished by the authors in Ref.~\cite{TsengWise}.
For comparison with our results, we just write down the power spectrum that they obtained
using the in-in formalism similar to the black hole case. The first order correction is~\cite{TsengWise}
\begin{eqnarray}
\Delta^{(1)}(\vec{k},\vec{q})&=& \frac{\epsilon
H^{2}\delta^{3}(\vec{k}+\vec{q})}{(2\pi)^{3}}\cdot\frac{1+kq\tau^{2}}{(k+q)kq}
\nonumber\\& & +\frac{\epsilon
H^{2}\delta(k_{z}+q_{z})}{(2\pi)^{5}}\cdot\frac{k^{2}+q^{2}+kq+k^{2}q^{2}\tau^{2}}{(k+q)k^{3}q^{3}}\times\nonumber\\
& &
\left[(2\pi)^{2}\delta(\vec{k}_{\perp}+\vec{q}_{\perp})\frac{\vec{k}_{\perp}\cdot\vec{q}_{\perp}}{2}
-2\pi\frac{\vec{k}_{\perp}\cdot\vec{q}_{\perp}}{(\vec{k}_{\perp}+\vec{q}_{\perp})^{2}}
+\frac{4\pi(\vec{k}_{\perp}+\vec{q}_{\perp})\cdot\vec{k}_{\perp}(\vec{k}_{\perp}+\vec{q}_{\perp})\cdot\vec{q}_{\perp}}
{(\vec{k}_{\perp}+\vec{q}_{\perp})^{4}}\right]
\,\,\,.\nonumber\\
\label{tsengstring}
\end{eqnarray}
In the result above the anisotropic part is the same as in Ref.~\cite{TsengWise} except for an overall sign. However, one can see that Eq.~(\ref{skspectrum2}) and Eq.~(\ref{tsengstring})
are still slightly different.

\subsection{Domain wall}

Using the metric~(\ref{dwall}) the Lagrangian density of a massless scalar field is given by
\begin{equation}
{\cal L}=-\frac{1-\epsilon^{2}}{2H^{2}\tau^{2}}\left[-(1-\epsilon^{2})(\partial_{\tau}\phi)^{2}+(\vec{\nabla}\phi)^{2}\pm2\epsilon(\partial_{\tau}\phi)(\partial_{z}\phi)\right].
\end{equation}
Then the conjugate momentum is
\begin{equation}
\Pi= \frac{\partial\mathcal{L}}{\partial\partial_{\tau}\phi}=
\frac{1-\epsilon^{2}}{H^{2}\tau^{2}}\left[(1-\epsilon^{2})\partial_{\tau}\phi\mp\epsilon\partial_{z}\phi\right],
\end{equation}
and the Hamiltonian is
\begin{eqnarray}
H&=&\int d^{3}x\left(\frac{1-\epsilon^{2}}{2H^{2}\tau^{2}}\right)\left[(1-\epsilon^{2})(\partial_{\tau}\phi)^{2}+(\vec{\nabla}\phi)^{2}\right]\nonumber\\
&=&\int d^{3}x\left(\frac{1-\epsilon^{2}}{2H^{2}\tau^{2}}\right)\left[\frac{H^4\tau^4\Pi^2}{(1-\epsilon^2)^3}+(\partial_{x}\phi)^{2}
+(\partial_{y}\phi)^{2}+\frac{(\partial_{z}\phi)^{2}}{1-\epsilon^2}\pm \frac{2\epsilon H^2\tau^2}{(1-\epsilon^2)^2}\Pi\partial_{z}\phi\right]\nonumber\\
&=&H_0+H_I.
\end{eqnarray}
To the linear order in $\epsilon$, we have the free Hamitonian and the interaction Hamitonian respectively,
\begin{eqnarray}
H_0&=&\int d^{3}x\left(\frac{1}{2H^{2}\tau^{2}}\right)\left[H^4\tau^4\Pi^2+(\vec{\nabla}\phi)^{2}\right],\\
H_I&=&\pm\int d^{3}x \,\epsilon\Pi\partial_{z}\phi.\label{HIdomain}
\end{eqnarray}

Following the same procedure as in the black hole and the cosmic string cases, we obtain
\begin{eqnarray}
\Delta^{(1)}(\vec{k},\vec{q})
&=&\frac{\epsilon H^{2}\tau(k_{z}q+q_{z}k)\delta(\vec{k}_{\perp}+\vec{q}_{\perp})}{16\pi^{4}k^{2}q^{2}(k_{z}+q_{z})}\nonumber\\
&&-\frac{i\epsilon H^{2}\tau^{2}(k_{z}q^{2}+q_{z}k^{2})\delta(\vec{k}_{\perp}+\vec{q}_{\perp})}
{32\pi^{4}k^{2}q^{2}(k_{z}+q_{z})}\times\nonumber\\
&&\ \ \ \ \ \ \ \left[\left(1-\frac{i}{k\tau}-\frac{i}{q\tau}-\frac{1}{kq\tau^{2}}\right){\rm Ei}[i(k+q)\tau]e^{-i(k+q)\tau}\right.\nonumber\\
&&\ \ \ \ \ \ \ \left.-\left(1+\frac{i}{k\tau}+\frac{i}{q\tau}-\frac{1}{kq\tau^{2}}\right){\rm Ei}[-i(k+q)\tau]e^{i(k+q)\tau}\right].
\end{eqnarray}
This is not the same as the one in Eq.~(\ref{wallpert}) obtained by the perturbation method. However, the structure of the two expressions are very similar.

If we take $\tau\rightarrow 0$, the correlator becomes
\begin{eqnarray}
\Delta^{(1)}(\vec{k},\vec{q})=\frac{\epsilon H^{2}(k_{z}q^{2}+q_{z}k^{2})\delta(\vec{k}_{\perp}+\vec{q}_{\perp})}{32\pi^{3}k^{3}q^{3}(k_{z}+q_{z})}.
\end{eqnarray}
This result has been derived recently in Ref.~\cite{solomonetal}.

\end{document}